\documentclass[10pt,final,twocolumn]{IEEEtran} 

\usepackage{ifthen}
\usepackage{graphicx}
\usepackage{amsmath}
\usepackage{amssymb}
\usepackage{setspace}
\usepackage{cite}

\newcommand{\imgwidth}{1} 
\newcommand{\bb}[1]{\mathbf{#1}}

\title{Optimal Real-Weighted Beamforming with Application to Linear and Spherical Arrays}
\author{V.~Tourbabin*,~\IEEEmembership{Student~Member,~IEEE},~M.~Agmon,~B.~Rafaely,~\IEEEmembership{Senior~Member,~IEEE} and J.~Tabrikian,~\IEEEmembership{Senior~Member,~IEEE}
\date{}
\thanks{

The authors are with the Department of Electrical and Computer Engineering, Ben-Gurion University of the Negev, Be'er-Sheva 84105, Israel (email: \{tourbabv,moraga,br,joseph\}@ee.bgu.ac.il)}}

\begin{document}
	\maketitle
\begin{abstract}
One of the uses of sensor arrays is for spatial filtering or beamforming. Current digital signal processing methods facilitate complex-weighted beamforming, providing flexibility in array design. Previous studies proposed the use of real-valued beamforming weights, which although reduce flexibility in design, may provide a range of benefits, e.g. simplified beamformer implementation or efficient beamforming algorithms. This paper presents a new method for the design of arrays with real-valued weights, that achieve maximum directivity, providing closed-form solution to array weights. The method is studied for linear and spherical arrays, where it is shown that rigid spherical arrays are particularly suitable for real-weight designs as they do not suffer from grating lobes, a dominant feature in linear arrays with real weights. A simulation study is presented for linear and spherical arrays, along with an experimental investigation, validating the theoretical developments.
\end{abstract}


\begin{center}
\textbf{EDICS Category: AUD-LMAP}
\end{center}

	\section{Introduction} 
\IEEEPARstart{S}{ignal} processing for microphone arrays of various configurations, including linear, planar and spherical, is widely studied and reported in the literature \cite{VanTrees, Krim, Williams, Rafaely2, Meyer2}.
One of the primary goals of microphone arrays is beamforming, or spatial filtering\cite{VanVeen}. 
It is one of the simplest methods for discriminating between different signals based on the physical location, distance, or direction of the sources. An effective way to achieve spatial filtering is using arrays of discrete sensors. When wired with independent conditioning electronics, discrete array sensors are flexible and are therefore very useful in time varying environments.

A variety of beamforming techniques is available, including data-independent, statistically optimal and adaptive approaches. Well studied examples are least squares beampattern synthesis\cite{VanTrees}, MVDR or general LCMV beamformers\cite{VanVeen} and the sidelobe canceller techniques\cite{Griffiths}. Although these beamformers are flexible and facilitate the design of arrays with desired spatial and frequency characteristics, they require complex array weights, i.e. with magnitude and phase specified at each frequency.

In order to achieve high spatial resolution, there is a need for a large number of channels resulting in very complicated beamformer structures. Usually, it is possible to simplify the structure by finding a compromise between performance and complexity\cite{Havelock1, Havelock2}. In this paper, we analyze the possibility to perform the processing using real-valued beamformer weights without the need for phase adjustments \cite{Agmon}. This approach appears in the adaptive antenna array processing literature. See for example the SOAC \cite{SAOC1,SAOC2,SAOC3}, RAMONA \cite{RAMONA1,RAMONA2,RAMONA3} and additional methods \cite{Choi,Choi1,Ghavami}. The main advantage of this approach is that, although complex weights offer greater control over the array response, using real weights requires only real arithmetic operations for the beamforming procedure. Therefore, the computational complexity can be considerably reduced. In \cite{SAOC1} a method for suppressing jammers by controlling the current amplitude only for a linear antenna array is described. It is shown that, although the total number of jammers is doubled with complex weights, using real weights results in a much faster computing speed and considerable hardware savings. 

In addition to efficient realization, real weights can also improve some aspects of array processing. 
With real weights, an improvement of the Gram-Schmidt orthogonal projection nulling algorithm \cite{Buehring} is achieved in \cite{RAMONA1}, where it is shown that the algorithm proposed there performs four times faster than the original one. Another paper by Choi \cite{Choi} presents an adaptive beamforming technique with real estimated weights. In this paper a statistical measurement model is assumed, which includes the desired signal, noise, and several interference sources. It is shown that the imaginary part of the measurements' covariance matrix is sufficient to span the interferer subspace. Thus, by using orthogonal subspace projection, a real-weighted beamformer can be calculated.
In \cite{Ghavami}, a data-independent real-weighted beamforming technique for a rectangular sensor array is described. A computationally efficient design algorithm is presented where the real array weights are obtained by constraining the beampattern to a desired response at several symmetrically-chosen points in frequency-angle space.   

In addition, the optimal real-weighted beamforming techniques can be utilized for designing beamformers implemented with continuous sensors like Electro-Mechanical Film (EMFI) \cite{Paajanen, Lekkala, Ealo1, Ealo2}, which have potential advantages over the traditional electret microphones, e.g. large aperture with a single sensor, although such a directional sensor may not be electronically steerable.

Although the above-mentioned publications presented real-valued beamforming techniques, there is a lack of system performance analysis based on common data-independent performance measures like directivity or sensitivity. Also, no beamforming technique has been described that achieves optimality with respect to these measures.
Furthermore, although it was evident from \cite{Ghavami} that real weights impose an undesired beampattern symmetry, yielding an additional parasitic main lobe and thus limiting the use of real-weighted beamformers in practice, this issue has not been systematically addressed in previous publications.

This paper presents novel formulation and analysis of real-weighted beamforming, offering the following contributions:
\begin{enumerate}
	\item Mathematical formulations are developed for optimal beamformers in terms of maximum beampattern directivity and
				minimum beamformer sensitivity subject to the constraint of using real weights. A closed-form solution for the
				optimal real weights is derived for general array geometry.
	\item The problem of beampattern symmetry induced by real weights is formulated, showing clearly the existence of an
				additional parasitic main lobe when used with open array geometries. This symmetry clearly degrades
				the flexibility and, in fact, the overall performance of open arrays with real weights.
	\item In order to overcome the symmetry imposed by real weights, it is shown that the undesired parasitic main lobe
				does not exist for rigid spherical arrays, making these arrays ideal for	real-weighted beamforming. The
				narrowband performance of spherical arrays that use real weights is shown to be comparable to the performance of
				spherical arrays that use complex-valued weights. It is also shown that the proposed method facilitates the
				design of spherical arrays with electronic steering capabilities.
\end{enumerate}

The optimal array processing problem described in this paper for the spherical array geometry can be regarded as a special case of the FIR based realization \cite{SunYan2} where only single weight filters are considered. This implies that the analytic solutions derived are comparable to the numerical solutions in the single-weight FIR realization.

The paper starts with a brief review of general beamforming and common performance measures. Next, the optimization problem is formulated for a general array geometry and the maximum directivity solution for real weights is derived. Simulation examples for linear and spherical arrays are then presented and the resulting beampattern properties are discussed. The results of directional sound field analysis \cite{Rafaely1} using the proposed optimal beampatterns, are presented for experimental data measured with spherical microphone array. Conclusions and future work follow.

	\section{Beamforming}
\label{beamf}
\subsection{Array Beampattern}
Consider an array consisting of $M$ sensors with an arbitrary geometry. 
Assuming a free field, the $q$'th sensor's output $y_q\left(\omega,\Omega,t\right)$ at a time $t$ as a response to a unit amplitude plane wave having frequency $\omega$ and arriving at direction $\Omega=\left(\theta,\phi\right)$, is given by:
\begin{equation}
	y_q\left(\omega,\Omega,t\right)=v_q\left(\omega,\Omega\right)e^{i\omega t},
	\label{eq:21a}
\end{equation}
where, $v_q(\omega,\Omega)$ is the gain and phase shift associated with location of the $q$'th sensor. The overall array output is given by:
\begin{align}
	\nonumber y\left(\omega,\Omega,t\right)&=\sum\limits_{q=1}^{M}{w_q\left(\omega\right)
	y_q\left(\omega,\Omega,t\right)}\\
	&=\sum\limits_{q=1}^{M}{w_q\left(\omega\right) v_q\left(\omega,\Omega\right)e^{i\omega t}},
	\label{eq:21b}
\end{align}
where, $w_q\left(\omega\right)$ is the weight applied by the beamformer to the $q$'th sensor at frequency $\omega$. The dependence on frequency will be omitted for notation simplicity. 
Now, by dropping the time dependence (as we are interested only in the response) and rewriting (\ref{eq:21b}) in matrix form, we get the beampattern of the array:
\begin{equation}
	B(\Omega)=\bb{w}^T\bb{v}(\Omega),
	\label{eq:21}
\end{equation}
where $\bb{w}=\left[w_1\,w_2\,...\,w_M\right]^T$ is the weights vector and $\bb{v}(\Omega)=\left[v_1\left(\Omega\right)\, v_2\left(\Omega\right)\, ...\,v_M\left(\Omega\right)\right]^T$ is the manifold or steering vector of the array. In general, the weights vector of an array incorporates its beamformer characteristics, while the manifold vector describes its geometry. 

Beamforming using real weights does not require phase shift implementation. Thus, from the mathematical point of view, we will simply assume that the weights vector is real-valued.
Later in this text we will discuss linear and spherical array geometries as examples of optimization techniques presented in Section \ref{optsol}.
\subsection{Performance Measures}
\label{permeas}
There are a number of performance measures for assessing the capabilities of a sensor array defined in the literature\cite{VanTrees}. The performance measures considered in this paper are: (i) Directivity, (ii) Sensitivity.

Directivity is defined as the ratio between the power of the output in the look direction to the average output power:
\begin{align}
	\nonumber D&=\frac{{|B(\Omega_l)|}^2}{\frac{1}{4\pi}\int\limits_0^{\pi}{\int\limits_0^{2\pi}{{|B(\Omega)|}^2\sin\theta
	d\phi d\theta}}}\\
	&=\frac{\bb{w}^T\bb{bb}^H\bb{w}^*}{\bb{w}^T\left(\frac{1}{4\pi}\int\limits_0^{\pi}
	{\int\limits_0^{2\pi}{{\bb{v}(\Omega)\bb{v}^H(\Omega)\sin\theta d\phi 
	d\theta}}}\right)\bb{w}^*},
	\label{eq:22}
\end{align}
where $\bb{b}=\bb{v}(\Omega_l)$ is the manifold vector in the look direction $\Omega_l$.
We define $\bb{C}$ as the following Hermitian matrix:
\begin{equation}
	\bb{C}\triangleq \frac{1}{4\pi}\int\limits_0^{\pi}{\int\limits_0^{2\pi}
	{{\bb{v}(\Omega)\bb{v}^H(\Omega)\sin\theta d\phi d\theta}}}.
	\label{eq:23}
\end{equation}
Finally, directivity can be written as:
\begin{equation}
	D=\frac{\bb{w}^T\bb{bb}^H\bb{w}^*}{\bb{w}^T\bb{C}\bb{w}^*}.
	\label{eq:24}
\end{equation}
For the case when the weights vector is real-valued we have:
\begin{equation}
	D=\frac{\bb{w}^T\bb{bb}^H\bb{w}}{\bb{w}^T\bb{C}\bb{w}}.
	\label{eq:24a}
\end{equation}

The second array performance measure discussed here is sensitivity, which is a measure of array robustness to processor gain errors, sensor mismatch, and imprecise positioning of the sensors. The sensitivity is proportional to the square of the weights vector norm \cite{VanTrees}:
\begin{equation}
	T_{se}={\|\bb{w}\|}^2=\bb{w}^H\bb{w}.
	\label{eq:25}
\end{equation}  
Once again, for the real-valued weights vector we simply obtain:
\begin{equation}
	T_{se}={\|\bb{w}\|}^2=\bb{w}^T\bb{w}.
	\label{eq:25a}
\end{equation}  

Needless to say, that it is always highly desirable to construct an array with the highest possible directivity and lowest possible sensitivity to errors. In the next section, we present the maximum directivity optimization problem and derive the maximum directivity real-valued beamformer subject to distortionless response constraint. We also discuss a lower bound on the sensitivity resulting from the distortionless response constraint and present a technique for finding an optimal directivity beamformer subjected to a constraint on the maximum sensitivity value.

	\section{Optimal Beamformers} 
\label{optsol}
\subsection{Complex-valued maximum directivity beamformer}
\label{mdcomp}
In this subsection, we present the formulation of the general optimization problem for the maximum directivity beamformer and its well known complex-valued solution. In the next subsection we will derive the real-valued solution for the maximum directivity optimization problem.

In order to maximize the directivity given by (\ref{eq:24}) subject to the distortionless constraint, i.e. $\bb{w}^T\bb{bb}^H\bb{w}^*=1$, it is sufficient to minimize $\bb{w}^T\bb{Cw}^*$. Note, that the distortionless response constraint used here is applied only to the absolute value of the response, leaving a degree of freedom in choosing the phase. The optimization problem can be stated as follows:
\begin{equation}
	\bb{w}_{c}=\arg\min\limits_{\bb{w}}\left(\bb{w}^T\bb{Cw}^*\right)\,\,\,\,\,\,\,\,s.t.\,\,\,\,\,\,\,\bb{w}^T\bb{bb}
	^H\bb{w}^*=1
	\label{eq:31a}
\end{equation}
The solution to this problem is known in the literature\cite{VanTrees, Rafaely6}, and is widely used to derive optimum beamformers such as MVDR or MPDR:
\begin{equation}
	\bb{w}_{c}=\left(\frac{\bb{C}^{-1}\bb{b}}{\bb{b}^H\bb{C}^{-1}\bb{b}}\right)^*.
	\label{eq:31b}
\end{equation}
However, this solution, in general, yields complex weights. A derivation of a real-valued solution, not available in literature, is presented in the following subsection.

\subsection{Real-valued maximum directivity beamformer}
\label{optsolA}
In this section, the real-valued solution to the maximum directivity beamformer optimization problem is derived. 
Assuming $\bb{w}$ is a real-valued vector and $\bb{C}$ is a Hermitian matrix, the problem can be formulated as follows:
\begin{equation}
	\bb{w}_{r}=\arg\min\limits_{\bb{w}}\left(\bb{w}^T\bb{Cw}\right)\,\,\,\,\,\,\,\,s.t.\,\,\,\,\,\,\,
	\bb{w}^T\bb{bb}^H\bb{w}=1
	\label{eq:31}
\end{equation}
Using the Lagrange multipliers method, the Lagrangian is given by:
\begin{equation}
	J(\bb{w},\lambda)=\bb{w}^T\bb{Cw}+\lambda\left(\bb{w}^T\bb{bb}^H\bb{w}-1\right),
	\label{eq:32}
\end{equation}
where $\lambda$ is the Lagrange multiplier. Taking the gradient with respect to $\bb{w}$ (assuming it is strictly a real-valued vector) and setting it to zero in order to find the minimum, we get:
\begin{equation}
	\bb{w}^T Re\left\{\bb{C}\right\}+\lambda\bb{w}^T Re\left\{\bb{bb}^H\right\}=0.
	\label{eq:33}
\end{equation}
We denote $\bb{\tilde{C}}=Re\left\{\bb{C}\right\}$. By right-multiplication of (\ref{eq:33}) with $\bb{\tilde{C}}^{-1}\bb{b}$, one obtains:
\begin{equation}
	\bb{w}^T\bb{b}+\lambda\bb{w}^T Re\left\{\bb{bb}^H\right\}\bb{\tilde{C}}^{-1}\bb{b}=0.
	\label{eq:34}
\end{equation}
Now the constraint is used to evaluate $\lambda$. According to the constraint $\bb{w}^T\bb{b}=e^{j\varphi}$, where $\varphi$ is an arbitrary real-valued phase. Thus (\ref{eq:34}) can be rewritten as:
\begin{equation}
	e^{j\varphi}+\lambda Re\left\{\bb{b}^H e^{j\varphi}\right\}\bb{\tilde{C}}^{-1}\bb{b}=0.	
	\label{eq:35}
\end{equation}
Now, $\lambda$ can be evaluated as
\begin{equation}
	\lambda=-\frac{1}{Re\left\{\bb{b} e^{-j\varphi}\right\}^T{\bb{\tilde{C}}}^{-1}\bb{b}e^{-j\varphi}}.
	\label{eq:36}
\end{equation}
By substituting (\ref{eq:36}) into (\ref{eq:33}) and simplifying, we get:
\begin{equation}
	\bb{w}_{opt}=\frac{\bb{\tilde{C}}^{-1}Re\left\{\bb{b} e^{-j\varphi}\right\}}{\left(\bb{b}
	e^{-j\varphi}\right)^T\bb{\tilde{C}}^{-1}Re\left\{\bb{b}
	e^{-j\varphi}\right\}}
	\label{eq:37}
\end{equation}
Finally, $\varphi$ can be found by forcing the imaginary part of (\ref{eq:37}) to zero. Since the numerator is real-valued, it is sufficient to set the imaginary part of the denominator to zero:
\begin{equation}
	\left[\bb{b}^T e^{-j\varphi}-\left(\bb{b}^T e^{-j\varphi}\right)^*\right]
	\bb{\tilde{C}}^{-1}Re\left\{\bb{b} e^{-j\varphi}\right\}=0.
	\label{eq:38}
\end{equation}
By expanding and simplifying (\ref{eq:38}), one obtains:
\begin{equation}
	\bb{b}^T\bb{\tilde{C}}^{-1}\bb{b}e^{-j2\varphi}=\bb{b}^H\bb{\tilde{C}}^{-1}\bb{b}^* e^{j2\varphi}.
	\label{eq:39}
\end{equation}
Solving (\ref{eq:39}) for $\varphi$, we get:
\begin{equation}
	\varphi=\frac{1}{2}\angle\left(\bb{b}^T\bb{\tilde{C}}^{-1}\bb{b}\right).
	\label{eq:310}
\end{equation}
Let $\bb{c} \triangleq Re\left\{\bb{b}e^{-j\varphi}\right\}$. Then optimal beamformer is given by:
\begin{equation}
	\bb{w}_{r}=\frac{\bb{\tilde{C}}^{-1}\bb{c}}{\bb{c}^T\bb{\tilde{C}}^{-1}\bb{c}}.
	\label{eq:311}
\end{equation}
It should be emphasized here that the matrix $\bb{C}$ defined in (\ref{eq:23}) includes the $\sin\theta$ factor. This factor can be treated as the cost function of the optimization problem. As we show later, other cost functions can be used as well, resulting in beneficial beampattern characteristics at the expense of degraded directivity. 

\subsection{Real-valued maximum directivity beamformer with bounded sensitivity}
In order to overcome possible errors and instabilities, a more robust beamforming technique is presented here. The following beamforming technique, which is based on the maximum directivity beamformer presented above, aims to find an optimal weights vector $\bb{w}$ that maximizes the directivity of a given array subject to bounded sensitivity, i.e. $T_{se}\leq T_0$. If the solution in (\ref{eq:311}) satisfies this constraint, it should be used directly. Otherwise, it is assumed that the constraint is active and is therefore satisfied with an equality, 
and the optimization problem with bounded sensitivity is formulated as follows:
\begin{equation}
	\bb{w}_{opt}=\arg\min\limits_{\bb{w}}\left(\bb{w}^T\bb{Cw}\right)\,\,\,\,s.t.\,\,\,\,\left\{
	\begin{array}{l}
		\bb{w}^T\bb{bb}^H\bb{w}=1\\
		\bb{w}^T\bb{w}=T_0
	\end{array}	
	\right.
	\label{eq:312}
\end{equation}
Using the Lagrange multipliers method, we write the Lagrangian as
\begin{align}
	\nonumber J(\bb{w},\lambda)=\bb{w}^T\bb{Cw}&+\lambda\left(\bb{w}^T\bb{bb}^H\bb{w}-1\right)\\
	&+\beta\left(\bb{w}^T\bb{w}-T_0\right).
	\label{eq:313}
\end{align}
By denoting $\bb{C}_d=\bb{C}+\beta \bb{I}$, where $\bb{I}$ is the identity matrix, the Lagrangian can be written as:
\begin{equation}
	J(\bb{w},\lambda)=\bb{w}^T\bb{C}_d\bb{w}+\lambda\left(\bb{w}^T\bb{bb}^H\bb{w}-1\right)-\beta T_0,
	\label{eq:314}
\end{equation}
Now, assuming $\beta$ is real-valued, one can proceed as in Section \ref{optsolA}. The optimal beamformer is given by:
\begin{equation}
	\bb{w}_{opt}=\frac{\bb{\tilde{C}}_d^{-1}\bb{c}}{\bb{c}^T\bb{\tilde{C}}_d^{-1}\bb{c}},
	\label{eq:315}
\end{equation}
where, $\bb{\tilde{C}}_d=Re\left\{\bb{C}+\beta\bb{I}\right\}$ and $\bb{c}=Re\left\{\bb{b}e^{-\frac{j}{2}\angle\left(\bb{b}^T \bb{\tilde{C}}_d\bb{b}\right)}\right\}$. The real-valued parameter $\beta$ can be found numerically by imposing the constraint $\bb{w}^T\bb{w}=T_0$.

It can be seen that the solution here is very similar to the maximum directivity beamformer. The only difference is the diagonal loading that appears in the matrix $\bb{\tilde{C}}_d$ which is expected to improve the robustness of the solution. 

\subsection{Lower bounds on sensitivity}
It is known that distortionless response constraint introduces a lower bound on the array's sensitivity\cite{VanTrees}. In the general case of complex-valued weights, the problem of minimum array sensitivity is mathematically identical to the problem in (\ref{eq:31a}) with the matrix $\bb{C}$ replaced by the identity matrix. Thus, the minimum sensitivity beamformer is given by:
\begin{equation}
	\bb{w}_{MS}^c=\frac{\bb{b}^*}{\bb{b}^H \bb{b}},
	\label{eq:315a}
\end{equation}
where the superscript $(\cdot)^c$ denotes the general case of complex-valued weights. Moreover, the minimum achievable sensitivity with complex weights is:
\begin{equation}
	T_{\min}^c={\bb{w}_{MS}^c}^H {\bb{w}_{MS}^c}=\frac{1}{\bb{b}^H \bb{b}}.
	\label{eq:315b}
\end{equation}
This is a well known result that reduces to $T_{\min}^c=\frac{1}{M}$ for open arrays, i.e. arrays with steering vectors consisting of phase shifting exponentials only. 

Note that the lower bound here depends strictly on $\bb{b}^H\bb{b}$, which is the only non-zero eigenvalue of the matrix $\bb{bb}^H$. We are going to exploit this observation later in order to establish a relationship between the lower bounds on sensitivity in complex and real-valued cases.

When the weights are constrained to be real-valued, the minimum sensitivity beamformer problem is, again, identical to (\ref{eq:31}) with $\bb{C}$ replaced by the identity matrix. The solution to this problem is given by (\ref{eq:311}):
\begin{equation}
	\bb{w}_{MS}^r=\frac{\bb{c}}{\bb{c}^T\bb{c}},
	\label{eq:315c}
\end{equation} 
where $\bb{c}=Re\{\bb{b} e^{-\frac{j}{2}\angle{\bb{b}^T\bb{b}}}\}$. Thus, the lower bound is given by:
\begin{equation}
	T_{min}^r={\bb{w}_{MS}^r}^T {\bb{w}_{MS}^r}=\frac{1}{\bb{c}^T \bb{c}}.
	\label{eq:315d}
\end{equation}

In order to get a more informative expression for $T_{min}^r$, we solve the minimum sensitivity problem directly, without using the solution in (\ref{eq:311}).
Consider the following minimization problem:
\begin{equation}
	\min\limits_{\bb{w}}\left(\bb{w}^T\bb{w}\right)\,\,\,\,\,\,\,\,s.t.\,\,\,\,\,\,\bb{w}^T\bb{bb}^H\bb{w}=1,
	\label{eq:316}
\end{equation}
where $T_{se}=\bb{w}^T\bb{w}$ is the sensitivity which we wish to minimize subject to the distortionless response constraint.
The Lagrangian will be:
\begin{equation}
	J(\bb{w},\beta)=\bb{w}^T\bb{w}-\beta\left(\bb{w}^T\bb{bb}^H\bb{w}-1\right).
	\label{eq:317}
\end{equation}
By taking the derivative with respect to $\bb{w}$, setting it to zero and rearranging, we get:
\begin{equation}
	Re\{\bb{bb}^H\}\bb{w}=\frac{1}{\beta}\bb{w}.
	\label{eq:318}
\end{equation}
Equation (\ref{eq:318}) holds for any pair of the scalar $\frac{1}{\beta}$ and the vector $\bb{w}$, which are an eigenvalue and eigenvector pair of the matrix $Re\{\bb{bb}^H\}$.

Note that $\bb{w}$ is a real-valued vector, hence it can be inserted inside the $Re\{\cdot\}$ operator. By multiplying both sides of (\ref{eq:318}) from the left by $\bb{w}^T$, and using the constraint in (\ref{eq:316}), we get:
\begin{equation}
	\bb{w}^T\bb{w}=\frac{1}{\gamma _i},
	\label{eq:319}
\end{equation}
where $\gamma _i$ is the $i^{th}$ eigenvalue of $Re\{\bb{bb}^H\}$. Equation (\ref{eq:319}) implies that for each $\bb{w}$, which is a solution of (\ref{eq:318}), there exists an appropriate eigenvalue of $Re\{\bb{bb}^H\}$ such that (\ref{eq:319}) holds.
It should be noted that the matrix $\bb{bb}^H$ has only one non zero eigenvalue, but by taking the real part, one can double the number of linearly independent rows. 
Thus, the minimum sensitivity is given by:
\begin{equation}
	T_{\min}^r=\min\limits_{\bb{w}}\left\{\bb{w}^T\bb{w}\right\}=\frac{1}{\gamma_{\max}},
	\label{eq:320}
\end{equation}
where, $\gamma_{\max}$ is the largest eigenvalue of matrix $Re\{\bb{bb}^H\}$. Note that the bound is achievable by the corresponding eigenvector of $Re\{\bb{bb}^H\}$, which in fact, is given by (\ref{eq:315c}).

In order to compare between $T_{\min}^c$ given in (\ref{eq:315b}) and $T_{\min}^r$, we realize that:
\begin{align}
	\nonumber \sum\limits_{i}{\gamma_i}&=\mathrm{tr}\left(Re\{\bb{bb}^H\}\right)=\mathrm{tr}(\bb{bb}^H)\\
	&=\bb{b}^H\bb{b}=\mu,
	\label{eq:321}
\end{align}
where $\mu$ is the non-zero eigenvalue of $\bb{bb}^H$. Hence, $\gamma_{\max}=\max\limits_{i}\{\gamma_i\}\leq\mu$ and
\begin{equation}
	T_{\min}^r\geq T_{\min}^c,
	\label{eq:322}
\end{equation} 
meaning that for a given array geometry and look direction, the real-valued beamformer can be at most as robust as the complex-valued one.

	\section{Linear Arrays}
\label{linarrgeom}
Here we demonstrate the maximum directivity beampattern that results from the theory developed above for a linear array with uniformly spaced sensors. In this case, the manifold vector is given by:
\begin{equation}
	\bb{v}(\psi)=\left[1\,\,e^{j\psi}\,\,e^{j2\psi}\,\,.\,.\,.\,\,e^{j(M-1)\psi} \right]^T,
	\label{eq:41}
\end{equation}
where $M$ is the number of sensors and $\psi=\frac{2\pi d}{\lambda}\cos\theta$ incorporates the spatial dependence indicating the phase difference between the signals in adjacent sensors as a function of plane wave arrival direction $\theta$. The angle $\theta$ is measured from the endfire side, and the origin of the coordinate system is positioned at the edge of the array. In addition, $\lambda$ is the wavelength of the impinging plane wave and $d$ is the spacing between the sensors.

The beampattern is given by:
\begin{equation}
	B(\psi)=\bb{w}^T\bb{v}(\psi)=\sum\limits_{n=0}^{M-1}{w_n\cdot e^{i\psi n}}.
	\label{eq:42}
\end{equation}
This form resembles the discrete-time Fourier transform (DTFT) implying that for real-valued weights vector $\bb{w}$, the beampattern will be symmetric in $\psi$ and thus in $\theta$. As a result, in addition to the main lobe at the desired look direction $\theta_l$, the beampattern will have a parasitic main lobe at $\pi-\theta_l$. 
The same is evident from the analysis performed in \cite{Ghavami}, although it is not stated explicitly. 

The beampattern of general three-dimensional array with real-valued weights is symmetric at least under the reversal of propagation direction. In order to see this consider an array consisting of $M$ microphones located at $\{\bb{r}_q\}_{q=1}^M$, where $\bb{r}_q$ is the position vector of $q^{th}$ microphone relative to a chosen coordinate system. For an open array, i.e. an array for which only the phase information is available, the response of $q^{th}$ sensor to a plane wave with the vawevector $\bb{k}_0$ is given by $e^{-j\bb{k}_{0}^T\bb{r}_q}$. Hence, the beampattern is given by:
\begin{equation}
	B(\bb{k}_0)=\sum\limits_{q=1}^{M}{w_{q}e^{-j\bb{k}_{0}^T\bb{r}_q}},
	\label{eq:42a}
\end{equation}
where $\{w_q\}$ are the real-valued weights. For the reversed propagation direction, i.e. $-\bb{k}_0$, we have:
\begin{align}
	\nonumber B(-\bb{k}_0)&=\sum\limits_{q=1}^{M}{w_{q}e^{j\bb{k}_{0}^T\bb{r}_q}}\\
	\nonumber &=\left(\sum\limits_{q=1}^{M}{w_{q}e^{-j\bb{k}_{0}^T\bb{r}_q}}\right)^*\\
	&=B^*(\bb{k}_0).
	\label{eq:42b}
\end{align}
Note, that for arrays with certain geometries, additional symmetries can arise as was shown above for the linear array. 
 
\begin{figure}[ht] 
	\centering
	\includegraphics[width=\imgwidth\columnwidth,keepaspectratio=true]{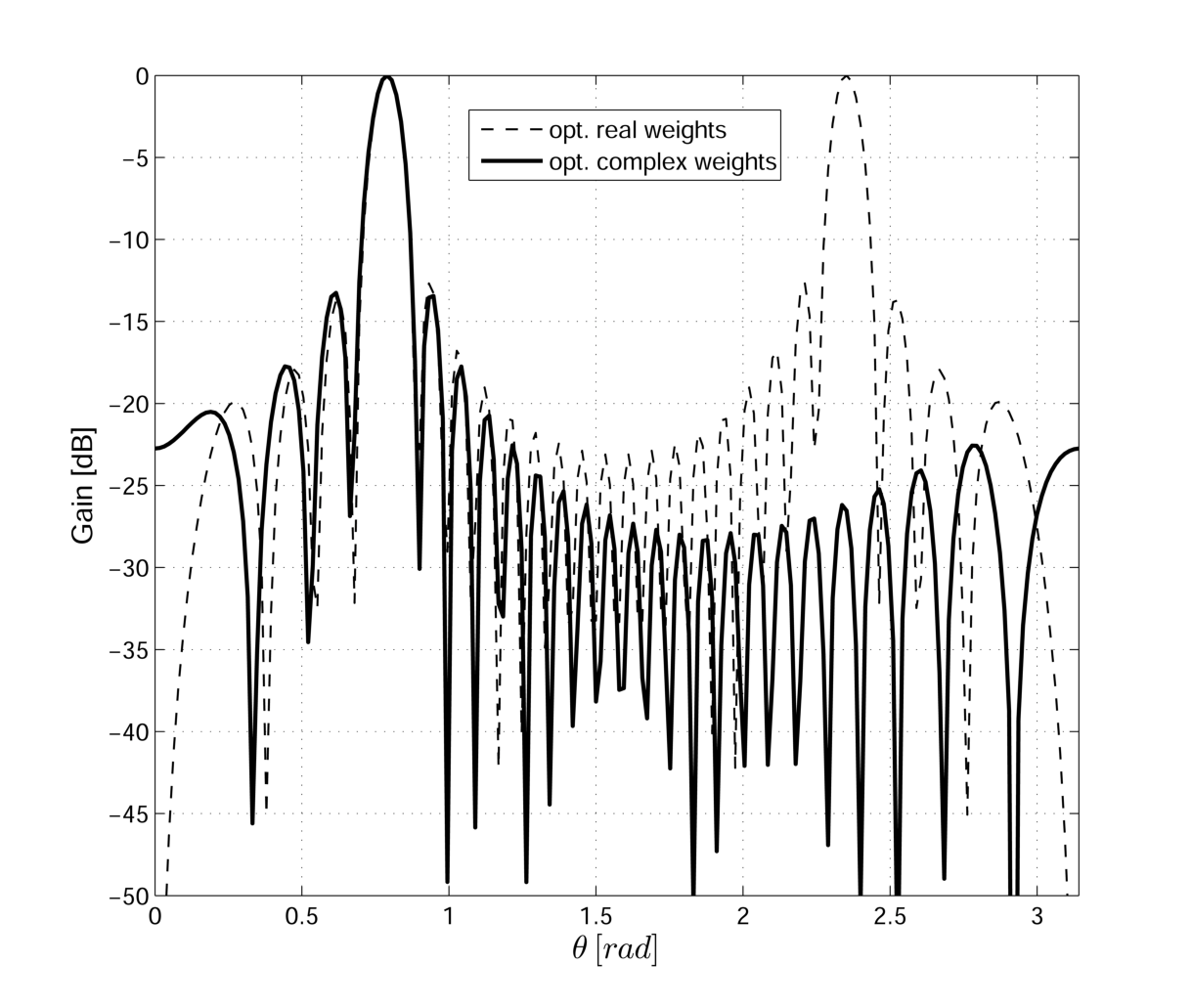}
	\caption{Comparison between two maximum directivity beampatterns - the real-weighted and the
	general complex-weighted, for a linear array with	parameters: $d=10$ cm, $M=25$, at $f=1715$ Hz.}
	\label{fig:41}
\end{figure}

Fig. \ref{fig:41} compares two maximum directivity beampatterns: the beampattern with real-valued weights derived above and the general beampattern with complex-valued weights \cite{VanTrees}. The beampatterns are presented at frequency $f=1715$ Hz for a uniform linear array consisting of $25$ sensors spaced at $10$ cm intervals (resulting in total array aperture of $2.4$ m). The frequency is chosen such that $\lambda=2d$ considering an acoustic medium with sound speed of $343$ m/s.

The array is steered to $\theta=\frac{\pi}{4}$ resulting in expected symmetry and a parasitic main lobe in the reciprocal direction $\theta= \frac{3\pi}{4}$ for the beamformer with real-valued weights. 
Ignoring the parasitic main lobe, there is a sidelobe with $-13$ dB level achieved for both beampatterns in this particular example. 

The resulting sensitivity of the beamformer with real-valued weights is $T_{se}=\bb{w}_{r}^T\bb{w}_{r}=0.076$, which is the same as the lower bound given by (\ref{eq:320}): $T_{\min}^r=0.076$. This result is not surprising. To see this, recall the matrix $\bb{C}$ defined in (\ref{eq:23}). For the linear geometry case the expression reduces to:
\begin{equation}
	\bb{C}=\frac{1}{2}\int\limits_{0}^{\pi}{\bb{v}(\theta)\bb{v}^H(\theta)\sin\theta d\theta},
	\label{eq:43}
\end{equation}
whose $nm^{th}$ element is given by: 
\begin{align}
	\nonumber [\bb{C}]_{nm}&=\frac{1}{2}\int\limits_{0}^{\pi}{e^{j\frac{2\pi d}
	{\lambda}(n-m)\cos\theta}\sin\theta d\theta}\\
	&=\mathrm{sinc}\left(\frac{2 d (n-m)}{\lambda}\right),
	\label{eq:44}
\end{align}
where $\mathrm{sinc}(x)=\frac{\sin\pi x}{\pi x}$. Thus, for the standard linear array, i.e. for $\lambda=2d$, as chosen above, $\bb{C}$ is the identity matrix. Hence, the problem of maximum directivity, formulated in (\ref{eq:31}), reduces to the minimum sensitivity problem and the solution is optimal in both senses. 

Generally, for the linear array geometry the matrix $\bb{C}$ will be close to the identity matrix for frequencies close to spatial Nyquist frequency of the array (i.e. $\lambda=2d$), meaning that the maximum directivity beamformer will be also nearly optimal in the sensitivity sense in this frequency range. However, for lower frequencies, there will be a strong deviation between the two optimal beamformers. These relationships will be demonstrated further for the spherical array geometry.

	\section{Spherical Arrays}
In this section, we apply the optimal beamformer derived above to spherical microphone arrays. These arrays are known for their advantages in processing three-dimensional sound fields due to their rotational symmetry. The spherical geometry enables the processing to be performed in the spherical harmonics domain\cite{Williams} also known as phase-mode processing\cite{Rafaely3}. 

\subsection{Phase-mode processing framework}
Here we will be concerned with sound fields which are scalar fields represented by a complex amplitude at each position in space. Consider the complex amplitude of the sound pressure $p(k,r,\Omega)$ on the surface of a rigid sphere of radius $r$ at angles $\Omega=(\theta,\phi)$, where $k=\frac{\omega}{c}$, with $\omega$ representing the angular frequency of the harmonic sound field and $c$ is the speed of sound. Spherical microphone arrays sample the surrounding acoustic field at positions $\{\Omega_i\}_{i=1}^M$, where $M$ is the total number of microphones. The output of the array is given by:
\begin{equation}
	y(k,r)=\sum\limits_{i=1}^{M}{\alpha_i p(k,r,\Omega_i) w^*(k,\Omega_i)},
	\label{eq:51}
\end{equation}
where $w^*$ is the weighting function of the array. The coefficients $\alpha_i$ are chosen in accordance with the specific sampling scheme \cite{Rafaely2}, allowing the expression of array output in terms of the spherical Fourier transform coefficients of the order-limited acoustic pressure and the weighting function:
\begin{equation}
	y(k,r)=\sum\limits_{n=0}^{N}{\sum\limits_{m=-n}^{n}{p_{nm}(k,r)w_{nm}^*(k)}}.
	\label{eq:52}
\end{equation}
The nearly-uniform sampling scheme will be used in the subsequent discussion, for which the coefficients are constant and given by $\alpha_i=\frac{4\pi}{M}$ \cite{Rafaely2}. 

Consider a unit amplitude plane wave $e^{i(\omega t+\bb{k}^T\bb{r})}$ impinging on a rigid sphere, where $\bb{r}=(r,\Omega)$ and $\bb{k}=(k,\Omega_0)$, with $\Omega_0$ indicating the arrival direction. The spherical Fourier transform coefficients of the resulting pressure on the sphere surface are given by:
\begin{equation}
	p_{nm}(k,r,\Omega_0)=b_n(kr){Y_n^m}^*(\Omega_0),
	\label{eq:53}
\end{equation}
where $Y_n^m$ are the spherical harmonics. The mode strength coefficients $b_n$ for a plane wave impinging on a rigid sphere are given by\cite{Meyer1}:
\begin{equation}
	b_n(kr)=4\pi i^n\left(j_n(kr)-\frac{j_n'(kr)}{h_n'(kr)}h_n(kr)\right),
	\label{eq:54}
\end{equation} 
where $j_n$ and $j_n'$ are the spherical Bessel functions of the first kind and their derivatives, respectively, and $h_n$ and $h_n'$ are the spherical Hankel functions of the second kind and their derivatives, respectively.

The spherical Fourier transform of a plane wave given in (\ref{eq:53}) is not order limited. But, in practice, the coefficients $b_n(kr)$ for $n>kr$ are negligible as compared to lower orders. Hence, spherical Fourier transform of a plane wave with a wavenumber $k<\frac{N}{r}$ can be represented by its lower order coefficients: $0\leq n\leq N$, with good accuracy.

Now, by substituting (\ref{eq:53}) for $0\leq n\leq N$ into (\ref{eq:52}) we get the array response to a plane wave arriving from the direction $\Omega_0$:
\begin{equation}
	y(k,r,\Omega_0)=\sum\limits_{n=0}^{N}{\sum\limits_{m=-n}^{n}{b_n(kr){Y_n^m}^*(\Omega_0)w_{nm}^*(k)}},
	\label{eq:55}
\end{equation}
which is recognized as the beampattern of the array. In the following, we will drop the frequency dependence for notation simplicity.

Here, we limit the discussion to weighting functions of the form:
\begin{equation}
	w_{nm}^*=d_n{Y_n^m}(\Omega_l),
	\label{eq:57}
\end{equation}
where $\Omega_l$ is the desired look direction. This particular form is highly useful resulting in an axisymmetric weighting function\cite{Rafaely2} with $\Omega_l$ indicating the axis of symmetry:
\begin{equation}
	w(\Theta)=\sum\limits_{n=0}^N{d_n\frac{2n+1}{4\pi}P_n(\cos\Theta)},
	\label{eq:58}
\end{equation}
where $\Theta$ is the angle between any direction $\Omega$ and the array look direction $\Omega_l$. The functions $\{P_n(\cdot)\}$ denote the Legendre polynomials which are real for real arguments. It can be seen that in order to have a real-valued weighting function it is sufficient that $d_n$ will be real.  

By substituting (\ref{eq:57}) into (\ref{eq:55}) we arrive at the desired expression for the beampattern of a spherical microphone array:
\begin{equation}
	B(\Theta)=\sum\limits_{n=0}^N{d_nb_n\frac{2n+1}{4\pi}P_n(\cos\Theta)}.
	\label{eq:59}
\end{equation}
It can be seen that the design parameters are the coefficients $d_n$, which can be chosen in an appropriate manner. After obtaining the coefficients, the actual microphone weights can be calculated by choosing the desired look direction and using (\ref{eq:57}) and (\ref{eq:58}). 

An important observation is that the phase-mode weights $d_n$ are designed independently from the look direction $\Omega_l$. As a consequence,  electronic steering of the beampattern can be accomplished by simple substitution of the desired look direction when the actual weights $w_{nm}$ are being calculated (see (\ref{eq:57})). 

We rewrite the expression in (\ref{eq:59}) using vector notation:
\begin{equation}
	B(\Theta)=\sum\limits_{n=0}^N{d_nv_n(\Theta)}=\bb{d}^T\bb{v}(\Theta),
	\label{eq:510}
\end{equation}
where $\bb{d}=[d_0\,\,d_1\,\,.\,.\,.\,\,d_N]^T$ and $\bb{v}=[v_0\,\,v_1\,\,.\,.\,.\,\,v_N]^T$, with $v_n=b_n\frac{2n+1}{4\pi}P_n(\cos\Theta)$.

It can be seen that the form of equation (\ref{eq:510}) is exactly the same as (\ref{eq:21}). The vector $\bb{d}$ can be considered as the weights vector in the spherical harmonics domain and $\bb{v}(\Theta)$ is again the manifold vector of the array.

Note that in the spherical case there is no undesired symmetry property similar to the symmetry apparent in linear arrays, because the functions $v_n(\Theta)$ in (\ref{eq:510}) are not symmetric in $\Theta$. This implies that the beampattern does not have a parasitic main lobe as in the case of linear and planar arrays where the symmetry is imposed by the conjugate symmetry of exponentials (see (\ref{eq:42})).

\subsection{Optimal phase-mode beamforming with real-valued weights}
Consider the phase-mode expression for the spherical array beampattern given in (\ref{eq:510}).
It is clear that by using the definition of directivity given in (\ref{eq:22}), one obtains the expression in (\ref{eq:24}) with $\bb{w}$ replaced by $\bb{d}$. Furthermore, the matrix $\bb{C}$ defined in (\ref{eq:23}) will be real-valued and diagonal as a straightforward result of the orthogonality property of the Legendre polynomials:
\begin{equation}
	\bb{C}=\left(\frac{1}{4\pi}\right)^2 \mathrm{diag}\left(|b_0|^2,3|b_1|^2,...(2N+1)|b_N|^2\right).
	\label{eq:511}
\end{equation}
Thus, one can directly apply the result in (\ref{eq:311}) in order to obtain the real-valued maximum directivity beamformer for the phase-mode processing of a spherical microphone array.

Spherical array sensitivity can be expressed using the phase-mode weights by starting from the expression for sensitivity in the spatial domain (see (\ref{eq:25}) and \cite{VanTrees}):
\begin{align}
	\nonumber T_{se}&=\sum\limits_{i=1}^{M}{|\alpha_i w(k,\Omega_i)|^2}\\
	\nonumber &=\frac{4\pi}{M}\sum\limits_{n=0}^{N}{\sum\limits_{m=-n}^{n}{|w_{nm}|^2}}\\
	\nonumber &=\frac{4\pi}{M}\sum\limits_{n=0}^{N}{|d_n|^2\sum\limits_{m=-n}^{n}{Y_n^m(\Omega_l){Y_n^m}^*(\Omega_l)}}\\
	\nonumber &=\frac{1}{M}\sum\limits_{n=0}^{N}{|d_n|^2 (2n+1)}\\
	&=\bb{d}^H\bb{U d},
	\label{eq:512}
\end{align}
where $\bb{U}=\frac{1}{M}\mathrm{diag}(1,3,5,...,2N+1)$. Note that the same expression, up to a normalization constant is used in \cite{SunYan1}, where the coefficients $\alpha_i$ are not considered as part of the array weighting vector.

The maximum directivity real-valued phase-mode beamformer with bounded sensitivity can be now obtained using the solution in (\ref{eq:315}), with matrix $\bb{\tilde{C}}_d=\bb{C}+\beta\bb{U}$. The $Re\{\cdot\}$ operator is omitted because both matrices $\bb{C}$ and $\bb{U}$ are real-valued for the spherical geometry case.

The lower bound on sensitivity and the minimum sensitivity beamformer will be given, again, by the solution of the optimization problem formulated in (\ref{eq:31}) with matrix $\bb{C}$ replaced by $\bb{U}$.

\section{Simulation Study}
\label{sphsim}
\subsection{Complex-valued vs. real-valued optimal beamformers}
\begin{figure} 
	\centering
	\includegraphics[width=\imgwidth\columnwidth,keepaspectratio=true]{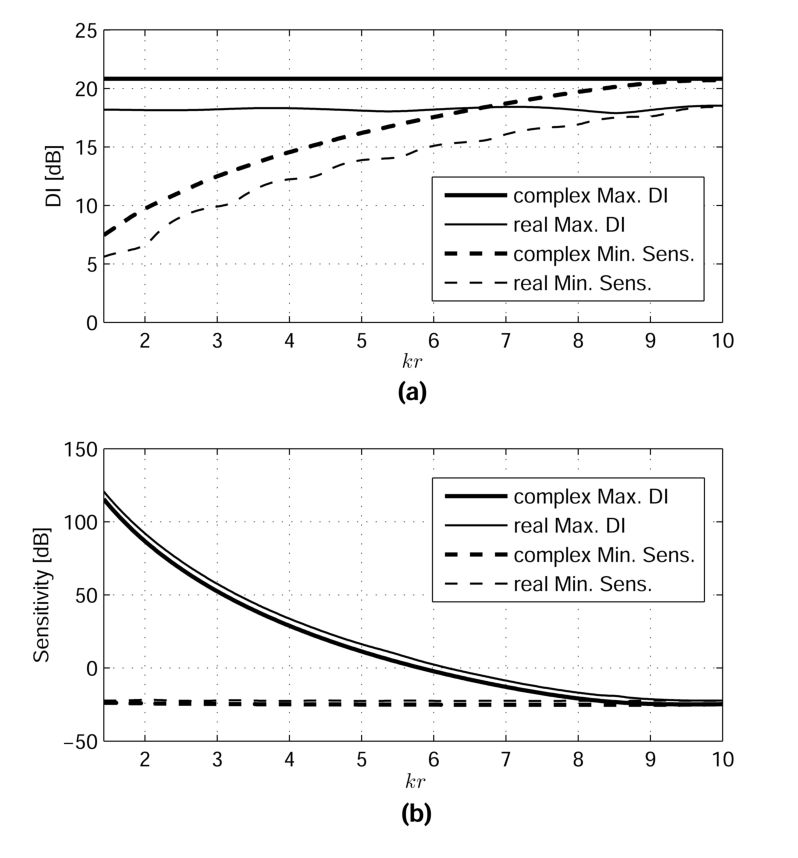}
	\caption{Comparison of the performance measures of complex-valued and real-valued optimal beamformers: (a) -
	directivity index, (b) - sensitivity.}
	\label{fig:50a}
\end{figure}

Recently, several publications focused on phase-mode beamforming methods with complex-valued weights for spherical microphone arrays \cite{Meyer1,Rafaely6,SunYan1,SunYan3,SunYan4,SunYan5}. The purpose of the current section is to compare the techniques with real-valued weights developed in this paper to their corresponding techniques with complex-valued weights \cite{Rafaely6}. The comparison is performed for spherical microphone array of order $N=10$. Fig. \ref{fig:50a} presents the performance measures of maximum directivity and minimum sensitivity beamformers as the function of the product $kr$. It should be noted that the minimum sensitivity beamformers were obtained using the solutions of the optimization problems in (\ref{eq:31a}) and (\ref{eq:31}) for the complex-valued and the real-valued cases, respectively, with matrix $\bb{C}$ replaced by $\bb{U}$ (see (\ref{eq:512})).

Recall that the real-valued solutions are sub-optimal in the sense that they result from the optimization over the space of real numbers which is a subspace of the complex numbers. Thus, real-valued beamformers are expected to perform poorer than the more general complex-valued beamformers. Nevertheless, it can be seen that the performance is not degraded significantly. The directivity indexes of the real-valued beamformers are lower than the directivity indexes of the complex-valued ones only by $3$ dB for the entire frequency range in this particular example. The difference in the sensitivities between the maximum directivity beamformers is about $5$ dB for lower frequencies, decreasing to $3$ dB for higher frequencies. 

In addition, note that in both cases (real and complex), the maximum directivity beamformers tend to be optimal also in the sensitivity sense at high $kr$ ranges. This behavior is similar to the one observed for the linear array geometry as discussed at the end of Section \ref{linarrgeom}. This is due to the fact that the absolute value of the mode strength coefficients $b_n(kr)$ when $kr$ approaches $N$ are nearly independent of the order $n$, i.e. $|b_n(kr)|\approx |b_0(kr)|$. This implies that $\bb{C}\approx \frac{M |b_0(kr)|^2}{(4\pi)^2}\bb{U}$. Hence, the maximum directivity and the minimum sensitivity problems tend to be the same for higher frequencies.

\subsection{Sidelobe performance}
\begin{figure}[ht] 
	\centering
	\includegraphics[width=\imgwidth\columnwidth,keepaspectratio=true]{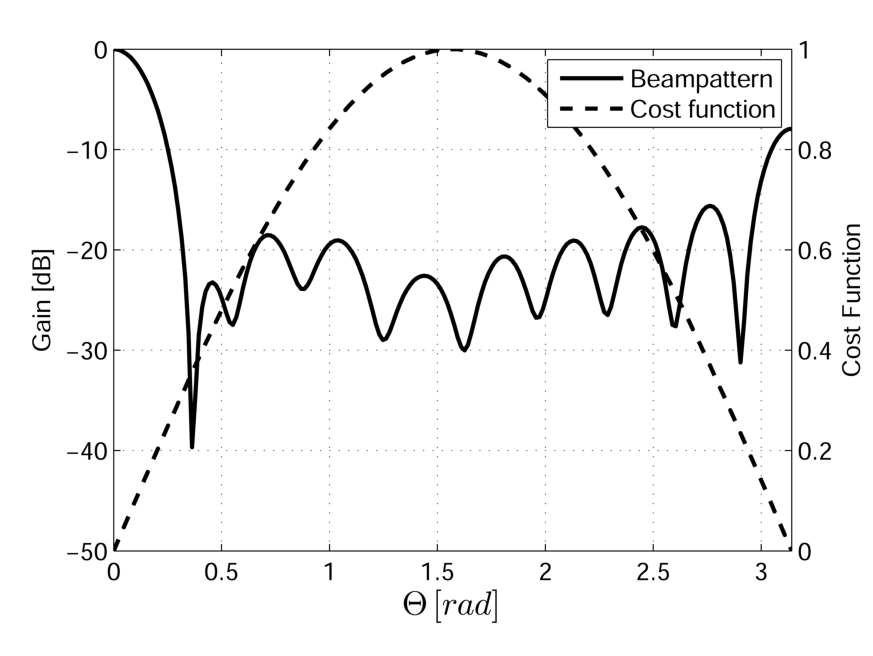}
	\caption{Maximum directivity beampatterns with real-valued weights for a spherical array of order $N=10$, at $kr=10$.}
	\label{fig:51}
\end{figure}

Here we use (\ref{eq:311}) and (\ref{eq:510}) in order to calculate the beampattern of the real-valued maximum directivity beamformer. Fig. \ref{fig:51} illustrates the simulated maximum directivity beampatterns at $kr=10$ for the spherical microphone array of order $N=10$. 

The array is steered to $\Theta=0$. It can be seen that there is indeed a main lobe in the desired direction with a sidelobe level of about $-8$ dB. As can be seen, the relatively poor sidelobe-level performance is due to a side lobe at $\Theta=\pi$.

In order to find the reason for the appearance of the sidelobe at $\Theta=\pi$ we return to the definition of directivity in Section \ref{permeas}. It was seen that matrix $\bb{C}$ defined in (\ref{eq:23}) includes the $\sin\Theta$ function (when the look direction $\Omega_l$ coincides with the positive $z$ axis, we get $\Theta=\theta$), which, as mentioned above, plays the role of a cost function. This implies that the solution to the optimization problem defined by (\ref{eq:31}) will tend to pull the energy towards $\sin\Theta=0$, minimizing the response at the center where the cost is high.

With this in mind, one can try different cost functions in order to affect the sidelobe level or the gain distribution along $\Theta$. Although cost functions other than $\sin\Theta$ will cause degradation in directivity, they can be beneficial in decreasing the sidelobe level, while at the same time affecting the main lobe width, as will be demonstrated shortly. 

\begin{figure}[ht] 
	\centering
	\begin{tabular}{c}
		\includegraphics[width=\imgwidth\columnwidth,keepaspectratio=true]{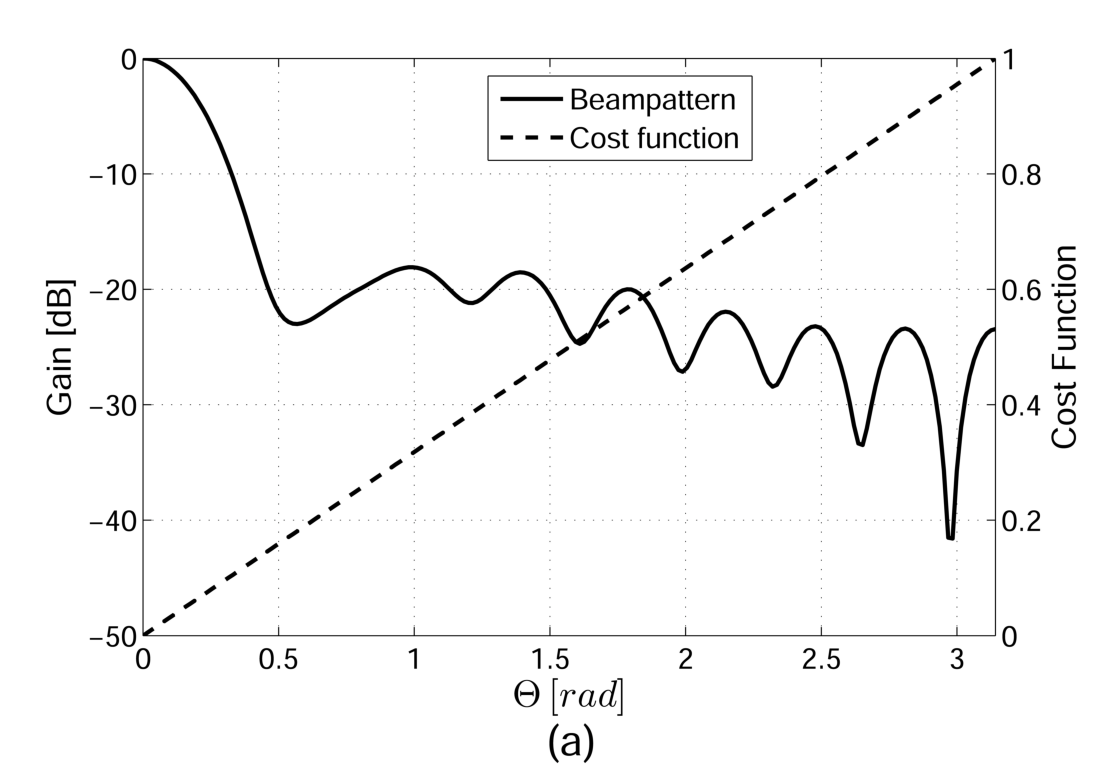}\\
		\includegraphics[width=\imgwidth\columnwidth,keepaspectratio=true]{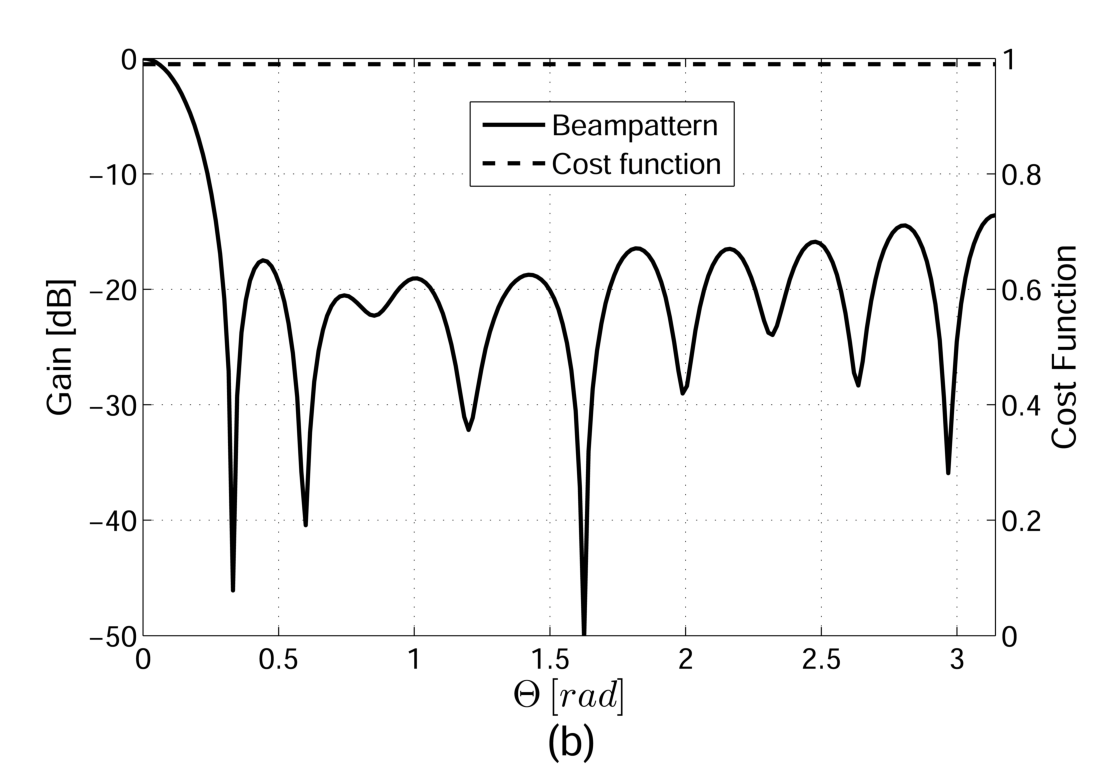}\\
		\includegraphics[width=\imgwidth\columnwidth,keepaspectratio=true]{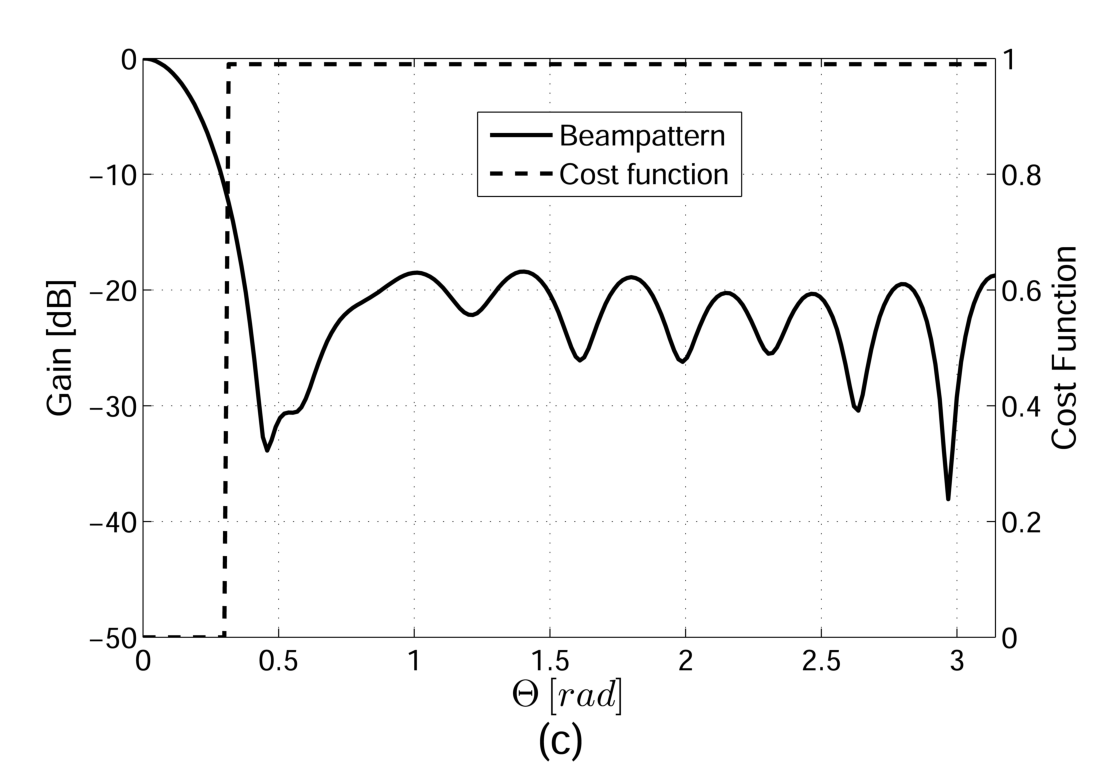}
	\end{tabular}
	\caption{Spherical array beam patterns using various cost functions: (a)-linear, (b)-uniform, (c)-step function.}
	\label{fig:52}
\end{figure}

Several alternative beampatterns were calculated using linear, uniform and unit step cost functions. The results are summarized in Fig. \ref{fig:52} and Table \ref{tab:51}.
It can be observed that the linear cost function (Fig. \ref{fig:52}(a)) imposes higher cost at $\Theta=\pi$ and the high-level sidelobe at $\Theta=\pi$ is avoided. In addition, using the unit step cost function (Fig. \ref{fig:52}(c)) and the uniform cost function (Fig. \ref{fig:52}(b)), results in a more uniform spread of side-lobe energy.  However, the three beampatterns in Fig. \ref{fig:52} have wider main lobes as compared to the maximum directivity beampattern (Fig. \ref{fig:51}) as well as lower directivity.  

Note that the array sensitivities using different cost functions are only slightly above the sensitivity of the maximum directivity beamformer (see Table \ref{tab:51}). Thus, at least for the parameters of the simulation presented here, the use of different cost functions did not impact on the array sensitivity. 
\begin{table} 
	\centering
	\caption{Comparison of beampattern parameters using various cost functions for spherical array with lower 
	sensitivity bound of $T_{\min}=-22.4$ [dB].}
	\begin{tabular}{lllll}
	\hline
	\bfseries Cost func. & \bfseries Sidelobe [dB] & \bfseries DI [dB] & \bfseries Sens. [dB]\\
	\hline
	sin&-7.9&18.5&-22.3\\
	linear &-18.1&17.3&-20.6\\
	uniform &-13.6&17.9&-21.8\\
	step function &-18.5&17.9&-21.3\\
	\hline
	\end{tabular}
	\label{tab:51}
\end{table}
	\section{Experimental Study}
Spherical microphone arrays can be utilized for sound field analysis. Examples of plane wave decomposition (PWD) of the sound field using maximum directivity complex-valued beamformer can be found in the literature \cite{Rafaely4}. However, the real-valued maximum directivity beamformer derived in this paper is suboptimal in the sense that the weights vector is constrained to be real-valued. 

Here, we compare the performance of maximum directivity beamformers, the optimal complex-valued beamformer and the suboptimal real-valued beamformer, by using experimental data.

The large-version of the Eigenmike\textsuperscript{\textregistered} spherical microphone array\cite{EigenmikeManual}, consisting of $32$ microphones mounted on a rigid sphere of radius $r=9$ cm was employed. Note that the array order is $N=4$, meaning that its maximum operating frequency is about $2400$ Hz. The measurements were performed in an anechoic chamber (anechoic from $300$ Hz) chamber with dimensions of $2\times2\times2$ m. A B\&K OmniSource 4295 loudspeaker was used. It was positioned such that the direction of propagation relative to the array was $(\theta,\phi)=(100^{\circ},160^{\circ})$. 
By selecting the response from the loudspeaker to the microphone array at a desired frequency, the vector $\bb{p}$ containing complex pressure amplitudes at different microphones is obtained. Then, the spherical Fourier transform coefficient $p_{nm}$ of the pressure on the sphere are calculated in the least squares sense using the pseudo-inverse of the transform matrix \cite{Rafaely5}. 

\begin{figure} 
	\centering
	\begin{tabular}{c}
		\includegraphics[width=\imgwidth\columnwidth,keepaspectratio=true]{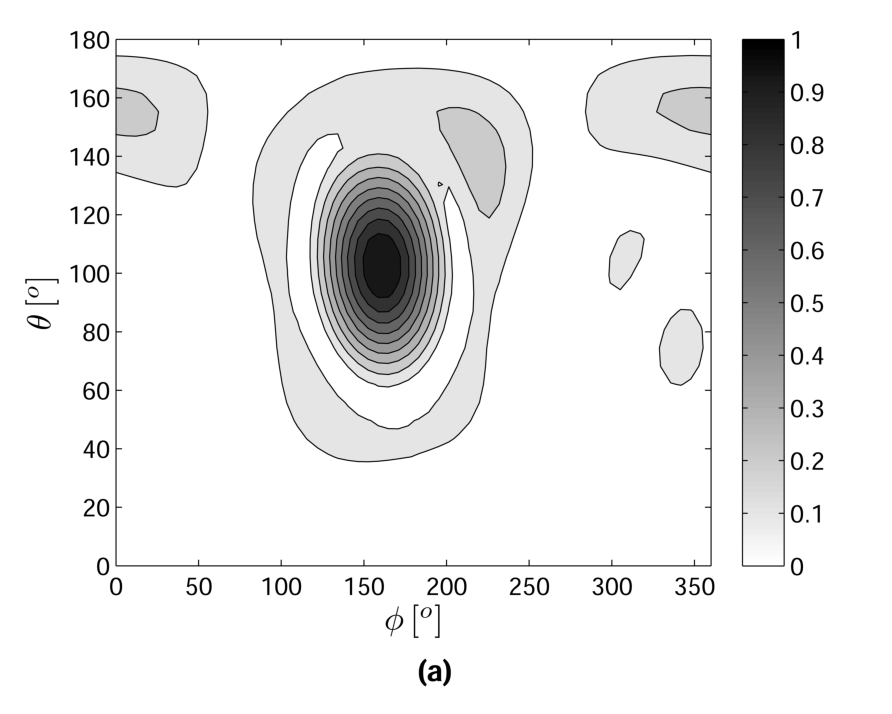}\\
		\includegraphics[width=\imgwidth\columnwidth,keepaspectratio=true]{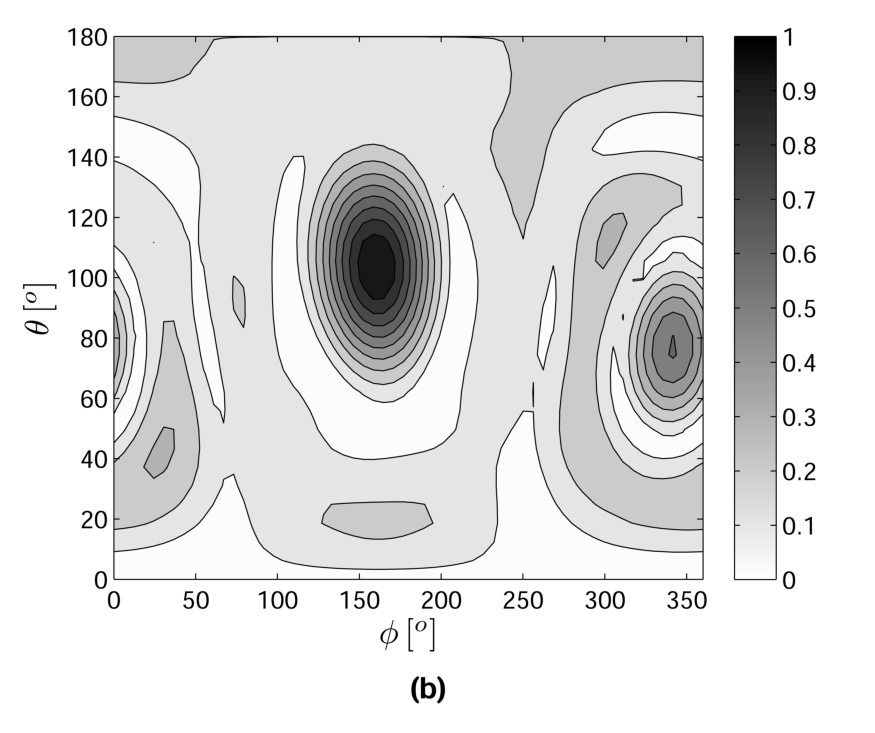}\\
		\includegraphics[width=\imgwidth\columnwidth,keepaspectratio=true]{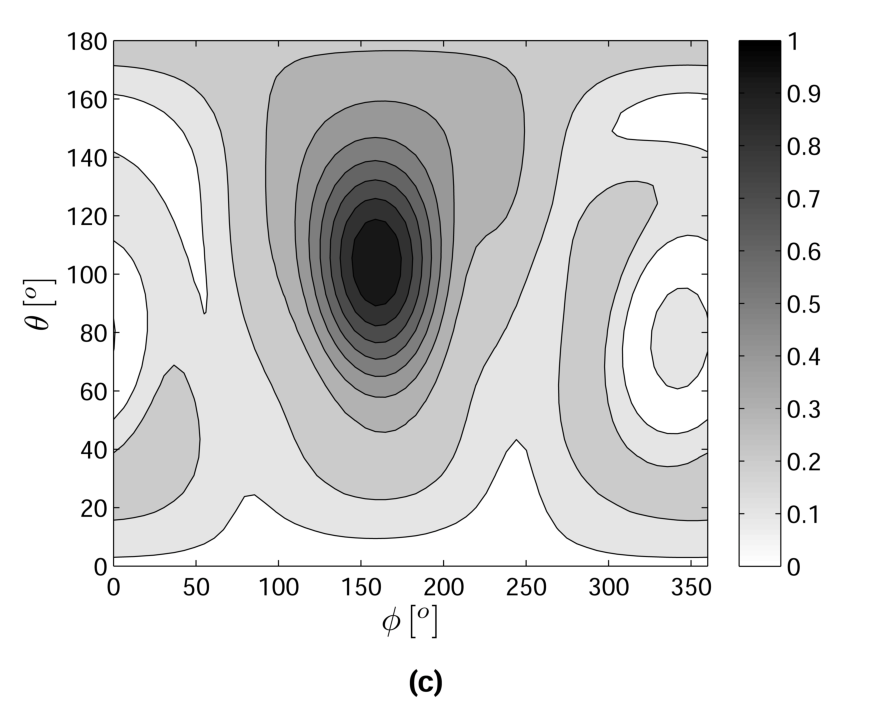}
	\end{tabular}
	\caption{Spatial sound-field analysis at $2400$ Hz ($N=4$, $r=9$ cm) using various beamformers: (a) complex-valued
	max. directivity (b) real-valued max. directivity (c)
	real-valued with linear cost function.}
	\label{fig:61}
\end{figure}

The spherical Fourier transform coefficients of the weighting function of the complex-valued maximum directivity beamformer are given by \cite{Rafaely2}:
\begin{equation}
	w_{nm}^*=\frac{Y_n^m(\Omega_l)}{b_n},
	\label{eq:61}
\end{equation}
meaning that $d_n=\frac{1}{b_n}$, which is equivalent to the result in (\ref{eq:31a}) up to a normalization constant $\frac{4\pi}{(N+1)^2}$ .
Thus, the output of the beamformer when steered to $\Omega_l$ is given by:
\begin{equation}
	y(\Omega_l)=\sum\limits_{n=0}^{N}{\sum\limits_{m=-n}^{n}{\frac{p_{nm}}{b_n}}Y_n^m(\Omega_l)}.
	\label{eq:62}
\end{equation}
Thus, PWD is performed by dividing the measured pressure coefficients by $b_n$ and transforming back into the spatial domain. Similarly, when using the optimal real-valued beamformer $\bb{d}=[d_0\,\,d_1\,\,.\,.\,.\,\,d_N]^T$ the pressure coefficients $p_{nm}$ are multiplied by $d_n$ and transformed into the spatial domain.
The results of spatial sound-field analysis using PWD are presented in Fig. \ref{fig:61}.
It can be seen that as expected, all three results have a maximum in the proximity of the loudspeaker direction. However, the analysis performed using real-valued maximum directivity beamformer (Fig. \ref{fig:61}.(b)) has an extra peak in the reciprocal direction. This is the straightforward consequence of the high sidelobe at $\Theta=\pi$ as discussed in Section \ref{sphsim}. It can be seen that, by using the real-valued beamformer with a linear cost function as proposed in section \ref{sphsim}, the undesired peak is completely eliminated. However, the resolution of the analysis is degraded as a consequence of the wider main lobe and lower directivity.

	\section{Conclusion}
Beamforming techniques with real-valued weights have the potential to significantly simplify beamformer structure.
Two optimal beamforming techniques using real weights were presented in this paper. It was shown that these techniques can be used with any desired array geometry. In particular, when used with spherical array geometry, the techniques result in electronically steerable beampatterns. A degree of flexibility was introduced by using various cost functions.
The results of numerical simulations and the experimental study show that, as expected, the performance of optimal beamformers with real-valued weights is lower than that of the optimal beamformers with complex-valued weights, but not considerably so in light of the potential hardware savings.

	\bibliographystyle{IEEEtran}
	\bibliography{IEEEabrv,allbib}
\begin{IEEEbiography} 
[{\includegraphics[width=1in,height=1.25in,clip,keepaspectratio]{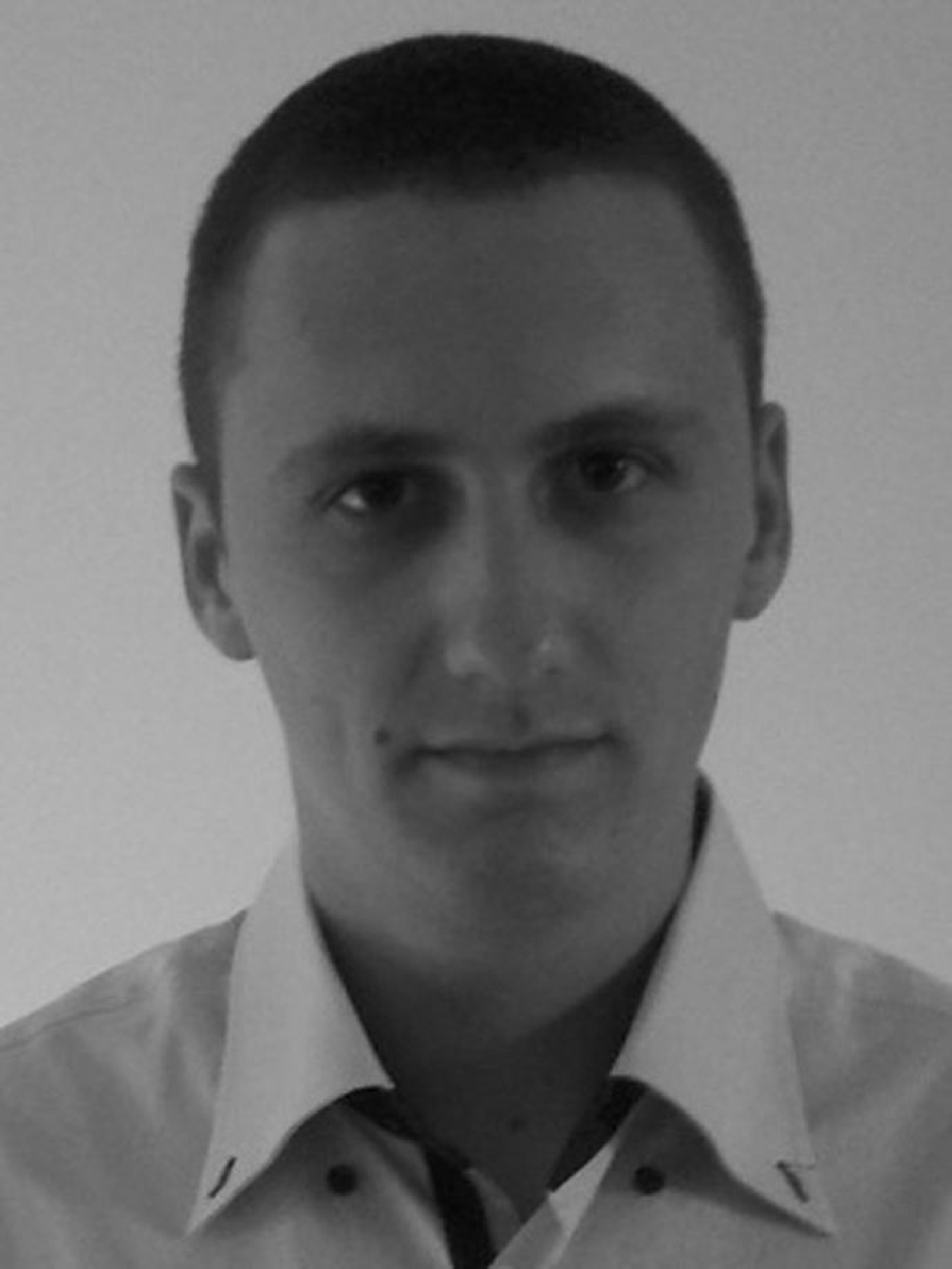}}]{Vladimir Tourbabin}
(S'12) received the B.Sc. degree (summa cum laude) in materials science and engineering and the M.Sc. degree (cum laude) in electrical and computer engineering from Ben-Gurion University of the Negev, Israel, in 2005 and 2011, respectively. He is currently working towards the Ph.D. degree in electrical and computer engineering at Ben-Gurion University.

His current research interests include microphone array signal processing and human sound perception. 

Mr. Tourbabin is a recipient of the Negev Faran Fellowship.
\end{IEEEbiography}

\begin{IEEEbiography} 
[{\includegraphics[width=1in,height=1.25in,clip,keepaspectratio]{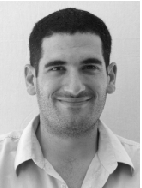}}]{Morag Agmon}
received the B.Sc. (cum laude) and M.Sc. degrees in electrical and computer engineering from the Ben-Gurion University of the Negev, Beer-Sheva, Israel, in 2008 and 2010, respectively.

He is currently leading the DSP Cellular Voice and Speech group in Marvell Semiconductor Inc. His research interests include acoustical signal processing and real-time microphone arrays.
\end{IEEEbiography}

\begin{IEEEbiography} 
[{\includegraphics[width=1in,height=1.25in,clip,keepaspectratio]{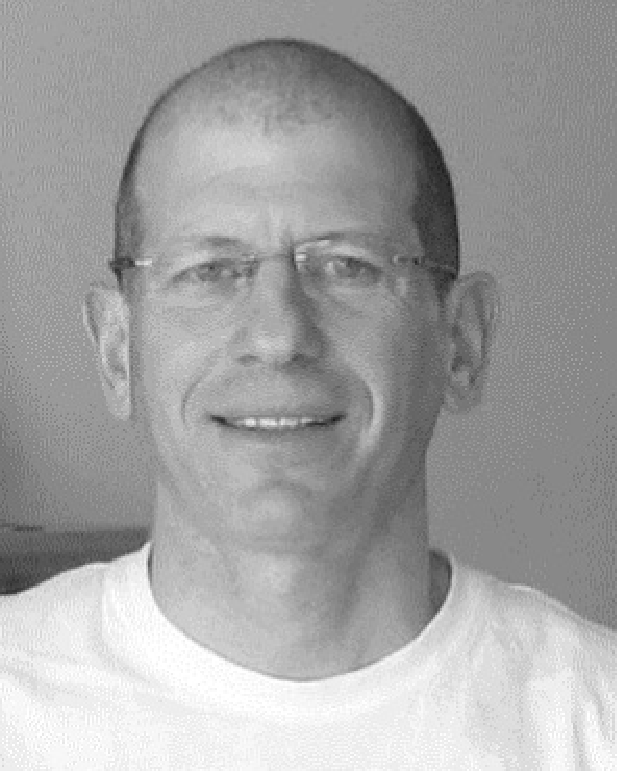}}]{Boaz Rafaely}
(SM'01) received the B.Sc. degree (cum laude) in electrical engineering from Ben-Gurion University, Beer-Sheva, Israel, in 1986; the M.Sc. degree in biomedical engineering from Tel-Aviv University, Israel, in 1994; and the Ph.D. degree from the Institute of Sound and Vibration Research (ISVR), Southampton University, U.K., in 1997. 

At the ISVR, he was appointed Lecturer in 1997 and Senior Lecturer in 2001, working on active control of sound and acoustic signal processing. In 2002, he spent six months as a Visiting Scientist at the Sensory Communication Group, Research Laboratory of Electronics, Massachusetts Institute of Technology (MIT), Cambridge, investigating speech enhancement for hearing aids. He then joined the Department of Electrical and Computer Engineering at Ben-Gurion University as a Senior Lecturer in 2003, and appointed Associate Professor in 2010. He is currently heading the acoustics laboratory, investigating sound fields by microphone and loudspeaker arrays. Since 2010, he is serving as an associate editor for IEEE Transactions on Audio, Speech and Language Processing. 

Prof. Rafaely was awarded the British Council's Clore Foundation Scholarship.
\end{IEEEbiography}

\begin{IEEEbiography} 
[{\includegraphics[width=1in,height=1.25in,clip,keepaspectratio]{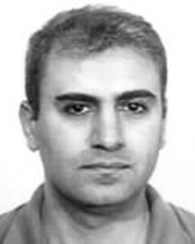}}]{Joseph Tabrikian}
(S'89-M'97-SM'98) received the B.Sc., M.Sc., and Ph.D. degrees in Electrical Engineering from the Tel-Aviv University, Tel-Aviv, Israel, in 1986, 1992, and 1997, respectively.

During 1996-1998 he was with the Dept. of ECE, Duke University, Durham, NC as an Assistant Research Professor. He is now with the Dept. of ECE, at the Ben-Gurion University of the Negev, Beer-Sheva, Israel. His research interests include estimation and detection theory and array signal processing. He has served as an Associate Editor of the IEEE Transactions on Signal Processing during 2001-2004, and since 2011, and as an Associate Editor of the IEEE Signal Processing Letters since 2012. He is a member of the IEEE SAM technical committee since 2010 and was the technical program co-chair of the IEEE SAM 2010 workshop. He is coauthor of 4 award-winning papers in different IEEE conferences.
\end{IEEEbiography}

\end{document}